\def\comment#1{}

\newcommand{\beg}{\begin{eqnarray}}
\newcommand{\eee}{\end{eqnarray}}

\documentclass[prl,letterpaper,twocolumn,aps,epsf]{revtex4}
\usepackage{dcolumn}
\usepackage{amsmath}
\usepackage{graphicx}%
\def\cm#1{}

\begin{document}
\title{
A superconductor to superfluid phase transition in liquid metallic hydrogen
}
\author{Egor Babaev${}^{1,2}$, Asle Sudb{\o}${}^2$, N. W. Ashcroft${}^1$}
\affiliation{
${}^1$Laboratory of Atomic and Solid State Physics, Cornell University,  Ithaca, NY 14853-2501 USA
\\ ${}^2$Department of Physics, Norwegian University of Science and Technology, N-7491 Trondheim, Norway 
}
\begin{abstract}
\end{abstract}
\maketitle
\newcommand{\la}{\label}
\newcommand{\aaa}{\frac{2 e}{\hbar c}}
\newcommand{\Pfaff}{{\rm\, Pfaff}}
\newcommand{\kA}{{\tilde A}}
\newcommand{\G}{{\cal G}}
\newcommand{\cP}{{\cal P}}
\newcommand{\M}{{\cal M}}
\newcommand{\E}{{\cal E}}
\newcommand{\btd}{{\bigtriangledown}}
\newcommand{\W}{{\cal W}}
\newcommand{\X}{{\cal X}}
\renewcommand{\O}{{\cal O}}
\renewcommand{\d}{{\rm\, d}}
\newcommand{\bfi}{{\bf i}}
\newcommand{\e}{{\rm\, e}}
\newcommand{\bfx}{{\bf \vec x}}
\newcommand{\bfn}{{ \vec{\bf  n}}}
\newcommand{\bfs}{{\vec{\bf s}}}
\newcommand{\bfE}{{\bf \vec E}}
\newcommand{\bfB}{{\bf \vec B}}
\newcommand{\bfv}{{\bf \vec v}}
\newcommand{\bfU}{{\bf \vec U}}
\newcommand{\bfp}{{\bf \vec p}}
\newcommand{\f}{\frac}
\newcommand{\bfA}{{\bf \vec A}}
\newcommand{\non}{\nonumber}
\newcommand{\be}{\begin{equation}}
\newcommand{\ee}{\end{equation}}
\newcommand{\ba}{\begin{eqnarray}}
\newcommand{\ea}{\end{eqnarry}}
\newcommand{\bastar}{\begin{eqnarray*}}
\newcommand{\eastar}{\end{eqnarray*}}
\newcommand{\half}{{1 \over 2}}

{ \bf
Although hydrogen is the simplest of atoms, it does not form the
simplest of solids or liquids. Quantum effects in these phases are
considerable (a consequence of the light proton mass) and they
have a demonstrable and often puzzling influence on many
physical properties  \cite {1}, including
spatial order. To date, the structure of dense hydrogen remains experimentally
elusive \cite{NWA2}. Recent 
studies of the melting curve of hydrogen \cite{Datchi2000,Bonev} 
indicate that at high (but experimentally accessible) pressures, compressed hydrogen
will adopt a liquid state, even at low temperatures. In reaching
this phase, hydrogen is also projected to pass through an
insulator-to-metal transition. This raises the possibility of new
state of matter: a near ground-state liquid metal, and its ordered
states in the quantum domain. Ordered quantum fluids are
traditionally categorized as superconductors or superfluids;
these respective systems feature dissipationless electrical
currents or mass flow. Here we report an analysis based on
topological arguments of
the projected phase of liquid metallic hydrogen, finding that it
may represent a new type of ordered quantum fluid. Specifically,
we show that liquid metallic hydrogen cannot be categorized
exclusively as a superconductor or superfluid. We predict that, in
the presence of a magnetic field, liquid metallic hydrogen will
exhibit several phase transitions to ordered states, ranging from
superconductors to superfluids.
%
}

Hydrogen constitutes more than 90\% of all atoms in the visible universe and
contributes three quarters of its mass. It is widely accepted that hydrogen is
abundant in the interiors of Saturn and Jupiter where it is both liquid and
metallic \cite{Science}, and the origin of their magnetospheres. The conditions in these
planets, particularly those of elevated temperatures, impel a view of dense
hydrogen as a classical liquid metal \cite{NeilPRA}. In what follows, a quite different view is taken
for low temperatures where, for a range of densities, hydrogen is projected to
take up a state which may be described as a quantum liquid metal. The notion
originates both with the light mass of the proton and the form of the
electronically screened, and hence density dependent, proton-proton
interactions.

The proton has one fourth the mass of ${}^4He$,
which in a condensed phase at normal conditions is  a classic {\it permanent liquid},
a consequence of high zero-point energy compared with relatively weak ordering energies
arising from interactions. Similarly, zero-point energies of protons in a dense environment
are also high, and at elevated compressions there is a shift of electron density from intra-molecular 
regions to inter-molecular, and with it a progressive decline in the effective inter-proton attractions 
(both within proton pairs, and between). Because of this there is also a decline of ordering energies from 
interactions relative to protonic zero-point energies, and arguments have therefore been advanced 
\cite{NWA2} first to suggest the occurence of a melting point maximum in compressed hydrogen, 
but second that there may also be a range of densities where, as in ${}^4 He$, hydrogen may choose
a fluid phase for its ground state. En route it passes through an insulator-metal
transition and the phase will aptly be described as  liquid metallic hydrogen, a 
translationally invariant two-component fermionic liquid. There is preliminary experimental evidence 
that a melting point maximum may indeed exist \cite{Datchi2000} and it has received recent theoretical 
backing \cite{Bonev}. Experimentally a 12.4 fold compression of hydrogen has already been achieved at 
around 320 GPa. Estimates suggest that LMH should appear at 13.6 fold compression at pressure in the 
vicinity of 400 GPa  \cite{Bonev},
whereas hydrogen alloys may exhibit metallicity at significantly lower pressures \cite{ch4}.
 A  predicted key feature of LMH at low temperature is 
the {\it coexistence of superconductivity of proton-proton and electron-electron 
Cooper pairs} \cite{NWA2000}. These condensates are independently conserved,
since electronic Coopers pairs cannot be converted to protonic Cooper pairs.
Thus, there is no intrinsic Josephson coupling between the two condensates.
This sets LMH apart from multi-component electronic condensates such as $MgB_2$. 
We therefore address some possible novel and
                         experimentally observable physics of this 
                         new state of matter. 
 {Our goal   is to discuss effects  
independent  of pairing mechanism or other microscopic details.
So in a search for {\it qualitatively} new
physics we base our  analysis solely on the {\it topology} of the 
proton-electron superconducting condensate.}

The free energy appropriate for LMH will be described by the following 
Ginzburg-Landau (GL) model
\beg
F& = &
 \frac{\left| \left( \nabla +
i e {\bf A}\right) \Psi_e \right|^2}{2 m_e} + 
\frac{\left| \left( \nabla -
i e {\bf A}\right) \Psi_p \right|^2}{2 m_p} 
\nonumber \\
& &+ { V} (|\Psi_{e,p}|^2)
+ \frac{{\bf B}^2}{2}
; ~~~{\bf B} = {\bf \nabla} \times {\bf A}.
\la{act}
\eee
the condensate order parameters are complex fields denoted by $\Psi_\alpha = |\Psi_\alpha|e^{i \phi_\alpha}$, 
where $\alpha=p,e$, with $p$ and $e$ referring to protonic and electronic Cooper pairs and 
${V} (|\Psi_\alpha|^2) 
=b_\alpha |\Psi_\alpha|^2+\frac{c_\alpha}{2}|\Psi_\alpha|^4+d|\Psi_p|^2|\Psi_e|^2$.
We have introduced the masses 
$m_e$ and $m_p$ of the electronic and protonic Cooper-pairs,
respectively, and $\pm e $ is the effective charge
of the Cooper-pairs in the two condensates. 
Apart from the Josephson term $\Psi_e^* \Psi_p + h.c.$ which is forbidden, 
as noted above, (\ref{act}) may include other terms  which merely introduce 
small quantitative changes to the effects discussed in this paper, and which 
thus may be omitted.
The GL free energy  can  be rewritten as \cite{fm,prl}
\beg
& & F= 
\f{1}{2}\f{\f{|\Psi_e|^2}{m_e}\f{|\Psi_p|^2}{m_p}}{
 \f{|\Psi_e|^2}{m_e} 
+\f{|\Psi_p|^2}{m_p} 
} (\nabla ({\phi_e+\phi_p}))^2 +
\nonumber \\ 
& &
 \f{2}{  \f{|\Psi_e|^2}{m_e} 
+\f{|\Psi_p|^2}{m_p} }
\Biggl\{\f{|\Psi_e|^2}{2m_e}\nabla \phi_e-
\f{|\Psi_p|^2}{2m_p}\nabla \phi_p  \nonumber \\
& & - e\Biggl( \f{|\Psi_e|^2}{m_e} 
+\f{|\Psi_p|^2}{m_p} \Biggr){\bf A}\Biggl\}^2+
 \f{{\bf B}^2}{2}.
\la{new01}
\eee
The first term is recognised as the kinetic term of Gross-Pitaevskii functional 
for liquid ${}^4He$, since no coupling to a vector potential $\bf A$ is present.  
This term may be thought of as describing an {\it electrically neutral} mode in 
the system, and is nothing but dissipationless {\it co-directed currents of electrons 
and protons carrying zero net charge}. The second term is equivalent to a gauge-invariant 
gradient term in an ordinary superconductor describing a charged mode in the system.

From the point of view of electronic pairing it has been suggested that at certain densities 
metallic hydrogen is  a type-II superconductor \cite{ja,ja2} i.e. magnetic flux may penetrate 
it in the form of vortices. When both protonic and electronic pairings occur the interesting 
physical question centers on whether there is a  vortex structure for both, and there are 
several distinct possibilities. The main features of the  ground state of vortex matter in this 
model  are: ({\it i}) If the vortices of both components share the same core, such a composite 
vortex is characterized by phase windings ($\Delta\phi_e=\pm2\pi$, $\Delta\phi_p=\mp2\pi$),  and 
then it has a finite energy per unit length, and carries one flux quantum \cite{prl}. Only these 
types of composite vortices can actually be induced by a magnetic field.  By a phase winding, we 
mean here the line integral of a phase gradient around a closed contour. In contrast, the vortices  
($\Delta\phi_e=\pm2\pi$, $\Delta\phi_p=0$) and  ($\Delta\phi_e=0$, $\Delta\phi_p=\pm 2\pi$) carry 
a fraction of flux quantum and then a logarithmically divergent energy per unit length \cite{prl}.
({\it ii}) In the absence of an external magnetic field, the phase transitions in (\ref{act}) are  
driven by a proliferation of thermally excited closed loops of vortices 
($\Delta\phi_e=2\pi$, $\Delta\phi_p=0$) and  ($\Delta\phi_e=0$, $\Delta\phi_p=2\pi$) at 
critical temperatures $T_c^e$ and $T_c^p$, respectively \cite{ns,sss}. 
We stress that in zero applied field the system is  {\it superconducting} at {\it all} temperatures
below $T_c^e$. 

Next, we point out that application of an external magnetic field can change the physical 
state of LMH dramatically and may result in a novel type of quantum fluid. We first
 consider the type-II regime. We emphasize that LMH should allow great
flexibility in changing  the GL parameter $\kappa$  both for protons
and electrons by varying the applied pressure and temperature \cite{ja,ja2}. 
In a superconductor with only one type of Cooper pair, a lattice of Abrikosov vortices 
melts in a first-order phase transition at a temperature $T_M(B)$ which decreases with 
increasing magnetic field \cite{ns}. The physical possibilities for LMH are
far richer, as we shall see. At zero temperature in an external field the system allows 
only composite vortices with $\phi_p=-\phi_e$ [for such a vortex the first term in 
(\ref{new01}) is zero]. However, because of thermal fluctuations, at $T\ne 0$ a vortex 
$(\Delta\phi_e=2\pi,\Delta\phi_p=-2\pi)$ can split  {\it locally } into two elementary 
vortices  $(\Delta\phi_e=2\pi,\Delta\phi_p=0)+(\Delta\phi_e=0,\Delta\phi_p=-2\pi)$
as shown in Fig. 1.
\begin{figure}[htb]
\centerline{\scalebox{0.15}{\rotatebox{0.0}{\includegraphics{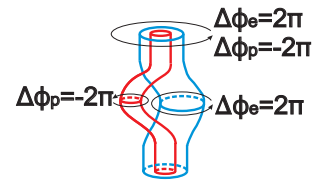}}}} 
\caption{Local splitting of a composite vortex line in LMH. The blue and red 
colors represent electronic and protonic vortices, respectively.
 The length scales are 
chosen almost equal for graphical convenience. }
 

\end{figure}
Such a splitting would result in a nontrivial contribution to the 
Ginzburg-Landau energy from the first term in (\ref{new01}), in the area 
in between two branches, since segments of such a loop attract each other 
logarithmically \cite{prl,sss}.
The system (\ref{act}) therefore possesses a characteristic ``{\it vortex ionization}" 
temperature  at which  a composite flux line $(\Delta\phi_e=2\pi,\Delta\phi_p=-2\pi)$ 
completely splits into two elementary vortices 
$(\Delta\phi_e=2\pi,\Delta\phi_p=0)+(\Delta\phi_e=0,\Delta\phi_p=-2\pi)$.
Such a splitting leads to a ``plasma" of line vortices interacting with  
a logarithmic potential.
This topological transition is in the $3D \ XY$ universality class, and should not 
be confused with topological phase transitions in two-dimensional superconductors.
In Fig.1 the protonic and electronic vortices     are represented by thin and thick lines
                        respectively. Of the two, the protonic  vortex fluctuates more because it 
                      has a {\it smaller} stiffness $|\Psi_p|^2/m_p$ due 
to {\it larger} mass $m_p\gg m_e$. It carries  a smaller fraction of flux quantum
$\Phi=\Phi_0|\Psi_p|^2/m_p[|\Psi_p|^2/m_p+|\Psi_e|^2/m_e]^{-1}$ \cite{prl}
and thus has smaller energy per unit length. This leads to 
a ``role inversion" in vortex matter: vortices in the condensate  of heavier particles cost less 
energy per unit length than vortices of condensates of lighter particles. 
  At low temperatures, the 
core size of a protonic vortex is expected to be much smaller than that of an electronic vortex 
because of the much larger   mass of the former. However such a picture is not applicable in the immediate vicinity 
of critical temperature for protons, where the protonic coherence length diverges.
We now proceed to discuss the LMH phase diagram at low and high magnetic fields.

{\bf LMH in low magnetic fields.}
Let us first consider the low-field regime in Fig. 2, when the characteristic temperature 
required to split a composite vortex line, $T_{SLM}$, is much smaller than the melting 
temperature of the lattice of  electronic vortices, $T_M^e$.  
Such a regime should be realizable in external fields much smaller than the 
upper critical magnetic field $H_{c2}^e$ for the electronic condensate.
Here, when the splitting of the field-induced  composite vortex line
$(\Delta\phi_e=2\pi,\Delta\phi_p=-2\pi) 
\to (\Delta\phi_e=2\pi,\Delta\phi_p=0)+(\Delta\phi_e=0,\Delta\phi_p=-2\pi)$
becomes of order of intervortex distances, the logarithmic interaction of the split 
vortices would be screened in a manner  similar to that expected in an ensemble of 
positively and negatively charged strings. 
When   $T < T_{SLM} \ll T_M^e$ we therefore have an Abrikosov lattice of composite 
vortices, which we may call a  {\it superconducting superfluid}  because of the 
coexistence of a neutral and charged modes, denoted {\bf SSF} in Fig. 2. However, 
upon   transition to the ``vortex-ionized"  state, $T > T_{SLM}$, we have {\it a 
lattice of electronic vortices immersed in a liquid of protonic vortex lines}.   There
has been {\it a vortex sublattice melting}, 
protonic superconductivity and the composite neutral mode disappear in this state, 
but the system  remains in an {\it electronic superconducting  
state} so long as the electronic vortex lattice remains intact,  i.e. for $T < T_M^e$ \cite{ns}.
This phase is denoted {\bf ESC} in Fig. 2; the phase transition separating
{\bf SSF} from {\bf ESC}, as well as the phase {\bf ESC} itself, have no 
counterparts so far in ordinary superconductors. 
One of the consequences of the presence of a background protonic vortex liquid is that 
electronic vortices  carry only  a fraction of the flux quantum, given by  \cite{prl}
$\Phi=\Phi_0 |\Psi_e(T)|^2/m_e[ |\Psi_e(T)|^2/m_e+ |\Psi_p(T)|^2/m_p]^{-1}$,
where $\Phi_0$ is the flux quantum. This fraction will be temperature dependent and with 
increasing temperature should reach the value $\Phi_0$ when $|\Psi_p|=0$. In addition to 
the  temperatures  $T_{SLM}$ and $T_M^e$, the system possesses characteristic 
temperatures $T_L^p(\Psi_p(B))$ and $T_L^e(\Psi_e(B))$ of  thermally driven  proliferation of protonic and 
electronic vortex loops, respectively, where $T_L^{p,e}(\Psi_{p,e}(B)) > T_{SLM}$. The zero-field
limit of $T_L^{p,e}(\Psi_{p,e}(B))$ corresponds to the temperatures  $T_c^{p,e}$ introduced below (\ref{new01}),
see Fig. 2. 
\begin{widetext}
\
\vskip -0.8cm
\begin{figure}[h]
\center{\includegraphics*[scale=0.065]{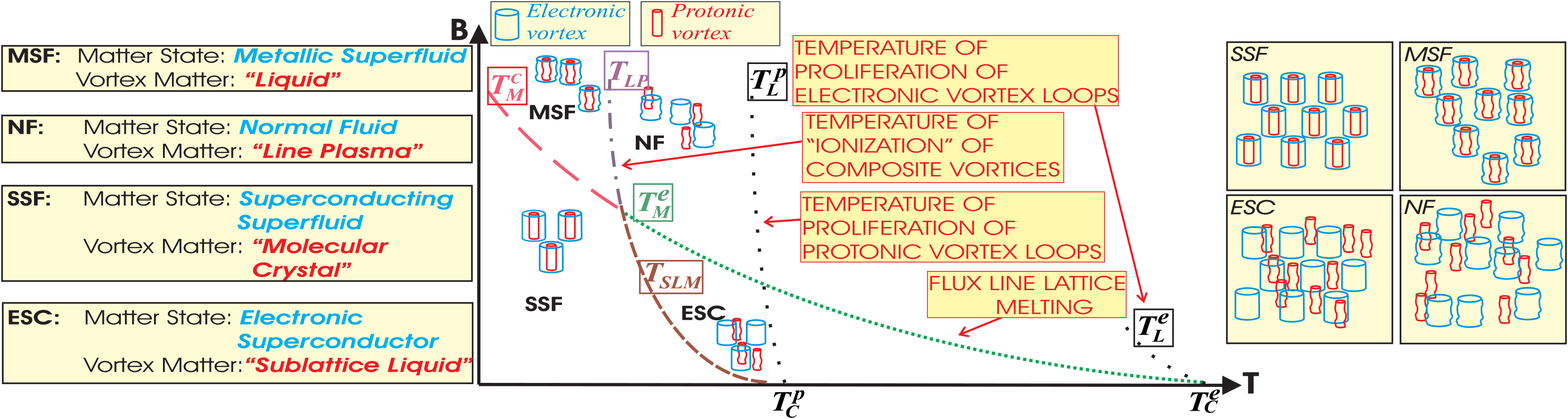}}
\caption{A schematic phase diagram of LMH as a function of applied magnetic field $\bf B$ 
and temperature $\bf T$. Phase {\bf SSF}: Composite vortex lattice, which is a superconducting 
superfluid state. Phase {\bf MSF}: Composite vortex liquid, 
which is a nonsuperconducting metallic superfluid state. {\it The transition from {\bf SSF} to 
{\bf MSF} is a superconductor-superfluid transition, and distinguishes LMH from any 
other known quantum fluid}. Phase {\bf ESC}: Electronic vortex lattice immersed in a 
protonic vortex line liquid. This is a superconducting, but not superfluid state. Phase 
{\bf NF}:  Vortex {\it line plasma}, which emerges when composite vortex lines are fully 
``ionized" into an electronic as well as a protonic vortices, neither of which is arranged 
in a lattice. 
It features nonzero resistivity as well as viscosity. 
The low-temperature vortex-liquid phase at very low magnetic
fields is not shown.
}
\end{figure} 
\
\vskip -1.1cm
\end{widetext}

{\bf LMH in strong magnetic fields.}
Here
the characteristic temperature required to split a composite 
vortex line, $T_{LP}$, is much larger
than the melting temperature of the lattice of  
composite vortices, $T_M^c$.  
Such a situation occurs when ({\it i}) the bare 
phase-stiffness of the electronic condensate $|\Psi_e|^2/m_e$ is 
suppressed by the external magnetic field down to being of the same 
order of magnitude as the  protonic stiffness,
 and ({\it ii}) the  characteristic 
temperature of  the melting of the lattice of composite vortices
is significantly lower than protonic and electronic critical temperatures 
(e.g. the electronic GL parameter $\kappa$ should be large, 
which should be achievable through choice of density \cite{ja,ja2}). 
The phase diagram then features the following hierarchy
of characteristic temperatures: 
({\it i}) $T_M^c$ - the melting temperature of the lattice of  composite 
vortices,  ({\it ii}) $T_{LP} > T_M^c$ - the ``vortex liquid" to ``vortex plasma"
transition temperature associated with 
fluxline splitting $(\Delta\phi_e=2\pi,\Delta\phi_p=-2\pi) \to
(\Delta\phi_e=2\pi,\Delta\phi_p=0)+(\Delta\phi_e=0,\Delta\phi_p=-2\pi)$.
As noted, this transition has a $3D \ XY$ universality class.
We emphasise that this transition is very different from the sublattice 
melting transition considered above.

Next, we examine the physical consequences of this hierarchy of  
characteristic temperatures. 
At low temperatures, the magnetic properties are controlled
solely by the charged mode, which is described by the second term
in (\ref{new01}). 
That is, at magnetic fields below the $T_M^c(B)$ line, 
the system exhibits a phase which is a field-induced
{\it lattice} of composite vortices $(\Delta\phi_e=2\pi,\Delta\phi_p=-2\pi)$
for which $\phi_p=-\phi_e$ and the first term
in (\ref{new01}) is exactly zero. This
corresponds to the {\it superconducting superfluid} discussed above.
However, increasing the magnetic      field to cross the $T_M^c(B)$
 line now leads to a first       order 
melting transition from a lattice to 
a liquid of composite vortices. This transition
is completely decoupled from the neutral superfluid
mode, while superconductivity is destroyed.
{\it It is therefore a first order phase transition from
a superconductor to a superfluid}. This  distinguishes 
LMH from any other known quantum fluid. It naturally requires
a revision of current classification schemes of quantum
fluids into the two categories of superconductors and 
superfluids. Indeed, the metallic superfluid state,
denoted  {\bf MSF} in Fig. 2, acquires all 
the attributes of superfluidity of neutral atoms like ${}^4 He$ 
even though microscopically it originates in a liquid  of {\it charged} 
Cooper pairs. 

We note that {\bf SSF} phase is characterized by
off-diagonal long-range order (ODLRO) in both fields
${\rm lim}_{|{\bf r}-{\bf r'}|\to \infty} <\Psi_\alpha({\bf r}) \Psi_\alpha^+({\bf r'})> \ne 0$
for $\alpha=p,e$.
In the {\bf MSF} state, the phases of both fields are disordered
and there is no  ODLRO for superconducting order parameters
(${\rm lim}_{|{\bf r}-{\bf r'}|\to \infty} <\Psi_\alpha({\bf r}) \Psi_\alpha^+({\bf r'})> = 0$).
In contrast, the  {\it neutral mode} retains ODLRO, manifested
by the preserved order in the phase sum $(\phi_p+\phi_e)$.
From this follows  a counterpart to the Onsager-Penrose 
criterion \cite{Onsager} { \it for metallic superfluidity:} 
${\rm lim}_{|{\bf r}-{\bf r'}|\to \infty} <\Psi_p({\bf r}) \Psi_e^+({\bf r'})> \ne 0$.
Under such circumstances,
the system is incapable of sustaining a dissipationless charge current, 
yet is capable of sustaining dissipationless massflow and consequently
 a vortex lattice induced
by {\it rotation}  as is possible in superfluid ${}^4 He$.
A rotation of a high-pressure diamond cell with hydrogen can be  
performed in an experiment in the presence of 
a cooling system, making such a rotating superfluid
state experimentally accessible in principle. An even more intriguing
possibility appears in case a rotation of  liquid metallic deuterium
since the deuteron has also spin degrees of freedom.
Increasing the temperature further 
produces an ``ionization" of  composite vortices. Eventually
superfluidity also disappears and we are left with a
metallic normal fluid; this corresponds
to the phase denoted by {\bf NF} in Fig. 2.

These observations should be of importance in
experimentally establishing that hydrogen may indeed take up a low temperature
liquid metallic state.
Experimental probes of the states of systems 
confined to high presure diamond cells are limited, but nonetheless application of external 
fields as well as the use of induction coils have already been successfully used to detect 
superconductivity at high pressures. The latter technique should also permit flux noise 
experiments.

Our  main points may be summarized as follows. ({\it i}) The vortex matter in LMH
is principally different from vortex matter in ordinary metallic superconductors. Our analysis 
shows that starting from a system of two types of fermions which form two distinct types 
of Cooper pairs, we arrive at what can be viewed as a {``dual condensed matter of 
vortices"}.
The vortices can be mapped onto a system of two types of charged strings which may be viewed 
as ``extended line particles"  with ``{\it reversed}" roles, namely the electronic vortices playing the role of  
``heavy particles" and protonic vortices being  ``light particles". Then, the Abrikosov lattice 
of composite vortices may be interpreted as a molecular crystalline state, which, at strong external 
fields  undergoes at $T_M^c$ a transition into a ``molecular liquid" and at higher temperature to a  
``plasma" state. In contrast, at weak external fields we find  a ``sublattice melting" transition, 
an intermediate state of {\it vortex} matter which has a counterpart in classical condensed matter physics 
as e.g. atomic sublattice melting in AgI. ({\it ii}) A particularly intriguing circumstance is that 
our analysis shows that an experimental  realization of LMH would mean that we have at hand a 
genuinely novel system which exhibits {\it a phase transition from a superconductor to a superfluid, 
or vice versa, driven by a magnetic field}. 
This  counterintuitive fact may require a revision of the standard classification
scheme of quantum liquids into superconductors and superfluids.


{\bf Acknowledgements} {\small 
This work was supported by the National Science Foundation, the Research 
Council of Norway, NANOMAT, and by STINT and the 
Swedish Research Council.}

{\bf Competing interests statement} {\small The authors declare that they have no competing financial
interests}.

{\bf Correspondence} and requests for materials should be addressed to E.B. (eb235 "(at)" cornell.edu).
\end{document}